# Bias factor of dislocation loops in quasicrystalline materials

Galina N. Lavrova[†], Anatoliy A. Turkin, Alexander S. Bakai

*Akhiezer Institute for Theoretical Physics, National Science Center "Kharkiv Institute of Physics & Technology", 1 Akademichna str. UA-61108 Kharkiv, Ukraine*

Vacancy swelling of quasicrystals under irradiation is considered. In quasicrystals, the evolution of dislocations is accompanied by the formation of phasons which are localized topological defects of the vacancy and interstitial types. At moderate temperatures the diffusivity of phasons is low which leads to the formation of ring-or disk-shaped phason trails inside dislocation loops. To find the capture efficiency of point defects by a dislocation loop with the complementary ring of phasons the steady-state drift-diffusional problem is solved in the toroidal geometry by the successive overrelaxation method (SOR). It is shown that phasons significantly reduce the bias of dislocations towards absorption of interstitial atoms. For this reason, quasicrystalline materials are predicted to exhibit increased resistance to vacancy swelling.

**Keywords**: quasicrystal; phasons; radiation effects in quasicrystals

**Introduction**

Irradiation of crystalline materials with energetic particles generates point defects (PD) – vacancies and interstitial atoms. The diffusion of PD to extended microstructural defects, such as dislocations, grain boundaries and voids, causes the evolution of material microstructure, which is observed as swelling, embrittlement and creep. In crystalline metals the majority of radiation effects are controlled by the dislocation bias towards absorption of interstitial atoms, as compared to vacancies. Unlike crystals with periodic lattices, quasicrystals lack translational symmetry. However, structural defects such as vacancies, interstitial atoms, dislocations and grain boundaries are inherent to quasicrystals, similar to crystalline materials. Phasons – the local rearrangements of atoms violating quasicrystalline order – are defects specific to quasicrystals [1]. It was found experimentally [2,3] that a dislocation climbing under applied stress, leaves behind a phason wall or trail. Similarly, in quasicrystals subjected to irradiation phasons are formed during the growth of dislocation loops due to dislocation bias toward absorption of interstitial atoms [4]. Phason defects of vacancy- and interstitial-types form clusters inside dislocation loops, which can gradually dissolve due to thermally activated phason diffusion. The activation energy for phason migration is rather high, i.e. up to 4 eV [5]. At moderate temperatures phasons remain inside the dislocation loop forming the phason trail, because thermally-activated diffusion of phasons is retarded [6].

---

[†] Corresponding author. e-mail: g.lavrova@kipt.kharkov.ua

In the previous paper the swelling behaviour of quasicrystals with mobile phasons [4], which are dispersed uniformly in the bulk of a quasicrystal, was studied. It was shown that radiation-induced vacancies and interstitials interact with phasons. After absorption of vacancies, phasons of the interstitial type transform into phasons of the vacancy type and vice versa. In other words, phasons are the recombination centers of alternating polarity for point defects. For this reason, the swelling rate of quasicrystals is lower as compared to that of crystals.

In this paper we consider a quasicrystal irradiated at temperatures when phasons remains inside the dislocation loop forming the phason trail. In a simplified model the phason trail can be viewed as a disc-shaped perfect sink for point defects. This assumption seems to be reasonable in the case of a high concentration of phasons when the inter-phason distances are short and a captured point defect can easily migrate to nearby phason defects prior to final absorption by the dislocation loop core.

In the next section we formulate the set of rate equations for PD in a material with dislocation loops and voids. The bias factor of dislocation loops is assumed to be the function of loop radius. Then the capture efficiencies of a dislocation loop for PD are calculated in a finite toroidal reservoir. Following Woo et al [7], the drift-diffusion problem of PD migrating in a stress field of the dislocation loop is solved numerically by a finite-difference technique known as the successive overrelaxation method (SOR) [8] based on Liebmann iterations [9]. Results of calculations allow one to evaluate (i) the effect of phason trail on the capture efficiencies of PD and (ii) to compare the swelling rate of quasicrystals with crystalline materials.

## Rate equations

In a quasicrystal under irradiation the behaviour of dislocations as a sink for PD differ from that in a crystal due to phason defects. It was experimentally established that a moving/climbing dislocation in a quasicrystal trails a phason fault which undergo gradual thermally activated "phason dispersion" [2]. In a similar way, in quasicrystals under irradiation the growth of dislocation loops is also accompanied by the formation of phason defects within each dislocation loop. In crystalline materials vacancies and interstitials created by irradiation recombine with each other and migrate to sinks. The rate equations for point defects in crystals are properly developed and studied [10]. Similar equations can be applied to describe the PD kinetics in irradiated quasicrystals taking into account the influence of phasons because they interact with vacancies and interstitial atoms. Phason defects interact with regular PD differently at low and high temperatures.

In this paper we consider the quasicrystal containing interstitial dislocation loops and voids under irradiation. It is known that the dislocation bias towards absorption of interstitial atoms results in accumulation of excess vacancies [11], which can form vacancy voids. We assume that thermally activated diffusion of phasons is retarded. In this case phasons remain localized within a dislocation loop in the form of a ring. When phasons are immobile, the rate equations for PD have the same form as those for crystals; however the sink strength of dislocation loops is modified by the presence of phason rings/disks inside loops. For simplicity we do not consider the nucleation stage of voids and dislocation loops. Thermal emission of vacancies from dislocation and voids is neglected.

With these assumptions, within the framework of the standard model of the effective medium [12] the rate equations for concentrations of vacancies $C_v$ and interstitial atoms $C_i$ are written as

$$\frac{dC_i}{dt} = K - \alpha_{iv} C_i C_v - k_i^2 D_i C_i, \tag{1}$$

$$\frac{dC_v}{dt} = K - \alpha_{iv} C_i C_v - k_v^2 D_v C_v, \tag{2}$$

where $K$ is the production rate of freely-migrating regular PD, $\alpha_{iv}$ is the recombination rate constants

$$\alpha_{iv} = 4\pi r_{iv}(D_i + D_v)/\Omega, \tag{3}$$

where $r_{iv}$ is the radius of spontaneous recombination of a vacancy with an interstitial atom and $\Omega$ is the atomic volume. The sink strengths $k_{i,v}^2$ consists of dislocation and void contributions

$$k_n^2 = k_{dn}^2 + k_C^2, \quad n = i, v \tag{4}$$

The sink strength of dislocations is given by

$$k_{dn}^2 = Z_{str,n}\rho_{str} + 2\pi \int Z_n(r) r f(r) dr \tag{5}$$

where $Z_{str,n}$ is the capture efficiencies of the straight dislocations, $Z_n(r)$ is the capture efficiencies of the dislocation loop of the radius $r$ and $f(r)$ is the size distribution function of dislocation loops normalised to the total number density of loops, $k_C^2$ is the void sink strength. In the following we assume that voids are unbiased sinks for vacancies and interstitials, i.e. the void sink strength is the same for vacancies and interstitial atoms

$$k_C^2 = 4\pi \int r F(r) dr, \tag{6}$$

where $F(r)$ is the size distribution function of vacancy voids normalised to the total void number density. Detailed calculations of the capture efficiencies of the dislocation loop with the phason ring/disk will be presented in the next section.

The growth rates of the dislocation loop of radius $r_L$ is defined by the difference of fluxes of interstitial atoms and vacancies [11]

$$\frac{dr_L}{dt} = \frac{1}{b}\left(Z_i D_i C_i - Z_v D_v C_v\right), \tag{7}$$

where $b$ is the Burgers vector. At a quasi steady state the defect concentrations are related by $k_v^2 D_v C_v = k_i^2 D_i C_i$. Therefore, the growth rate of the dislocation loop can be represented in the form

$$\frac{dr_L}{dt} = \frac{1}{b}\left[Z_i(r_L) D_i C_i - Z_v(r_L) D_v C_v\right] = \frac{1}{b}\left(B_L(r_L) - B_{mean}\right) Z_i(r_L) D_v C_v, \tag{8}$$

where $B_L = 1 - Z_v/Z_i$ is the bias factor of a loop of radius $r_L$ and $B_{mean}$ is the mean bias of all sinks in the system.

$$B_{mean} = \frac{k_i^2 - k_v^2}{k_i^2} = \frac{1}{k_i^2}\left(B_{str}Z_{str,i}\rho_{str} + 2\pi\int B_L(r)rf(r)Z_i(r)dr\right). \tag{9}$$

The void growth rate is proportional to the mean bias

$$\frac{dR_V}{dt} = \frac{1}{R_V}(D_vC_v - D_iC_i) = \frac{D_vC_v}{R_V}B_{mean}, \tag{10}$$

where $R_V$ is the void radius.

The swelling rate is represented as a growth rate of volume of all voids per unit volume of the material

$$\frac{dS}{dt} = \frac{d}{dt}\left(\frac{\Delta V}{V}\right) = k_C^2 D_v C_v B_{mean} \tag{11}$$

The quasi-steady-state concentrations of vacancies can be found from Equations (1) and (2)

$$C_v = \frac{D_i k_i^2}{2\alpha_{iv}}\left(\sqrt{1 + \frac{4K\alpha_{iv}}{D_i D_v k_i^2 k_v^2}} - 1\right) \tag{12}$$

At sufficiently high temperature when the bulk recombination of PD can be neglected the swelling rate is given by

$$S' = \frac{dS}{d\phi} = \eta B_{mean}\frac{k_C^2}{k_v^2}, \tag{13}$$

where $\phi = K_0 t$ is the irradiation dose according to NRT standard [13], $K_0$ is the dose rate according to NRT standard [13], $\eta$ is the fraction of freely-migrating regular PD [14].

Equations (8)-(12) show that the dependence of the bias factor on loop size significantly affects the evolution of an ensemble of loops and pores. The average bias factor is an important characteristic of the material microstructure, which controls the swelling rate of the material.

## Capture efficiencies of dislocation loops

In this section, we calculate capture efficiencies of interstitial dislocation loops for vacancies and interstitials in a quasicrystal. The migration of point defects into a dislocation loop surrounded by a sink-free region of influence, which is embedded into the effective medium, is considered. The effective medium described by Equations (1) and (2) plays a role of a reservoir maintained at constant average concentrations of point defects $\overline{C}$. Here a point defect is modelled as a dilation centre [15] with the relaxation volume $\Omega = \Omega_{v,i}$, different for vacancies and interstitials. The external boundary of the region of influence is selected in the form of a toroidal surface coaxial with the internal boundary which is the capture surface bounding the dislocation core. Here we consider the growing interstitial dislocation loop with the phason trail inside the loop. In order to estimate the maximum effect of phasons on capture efficiencies the phason trail is assumed to form a ring of the width $d$ which is a sink for point defects similar to dislocation core (Figure 1).

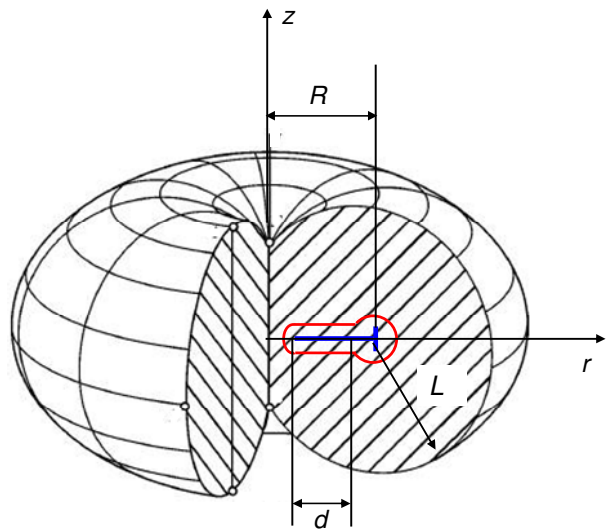

Figure 1. Sink-free region of influence of an interstitial dislocation loop of radius $R$. The small torus radius $L$ is of the order of the mean distance between sinks, $d$ is the width of the phason ring.

For numerical calculations the toroidal configuration is convenient for any size of a dislocation loop. At $R \ll L$, i.e. for an infinitesimal loop, the region of influence is close to a sphere; at $R \gg L$ the region of influence transforms into a cylinder enclosing the straight dislocation line [16]. Because of the rotational symmetry about the z-axis the corresponding boundary value problem can be formulated as a 2D problem (Figure 2).

The steady state diffusion-drift equation is given by

$$div\mathbf{j} = 0, \qquad \mathbf{j} = -D\nabla C - \frac{DC}{k_B T}\nabla U, \qquad (14)$$

where **j** is the flux of point defects, $D$ is the diffusion coefficient, $k_B$ is the Boltzmann constant, $T$ is the temperature, $U$ is the interaction energy of a point defect with the elastic field of the dislocation loop [17],

$$U(r,z) = -\frac{\mu\Omega(1+\nu)}{3\pi(1-\nu)} \frac{b}{\sqrt{(R+r)^2+z^2}} \left[ K(k) + E(k)\frac{R^2-r^2-z^2}{(R-r)^2+z^2} \right], \quad k^2 = \frac{4rR}{(R+r)^2+z^2}, \quad (15)$$

where $\mu$ is the shear modulus, $b$ is the Burgers vector, $\Omega$ is the relaxation volume, $\nu$ is the Poisson ratio, $K(k)$ and $E(k)$ are complete elliptic integrals of the first and second kind, respectively.

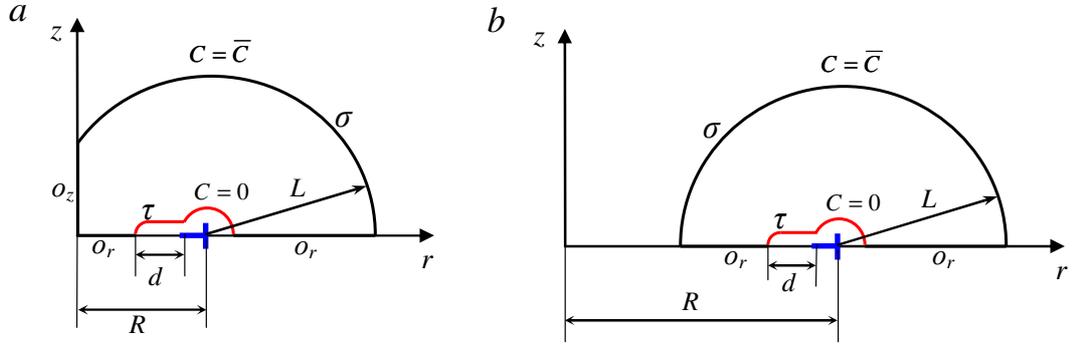

Figure 2. *Disposition of the co-ordinate system used for toroidal reservoir: (a) $R < L$ and (b) $R > L$.*

At the internal boundary of the sink-free region, at the dislocation core and phason ring, the concentration of PD is assumed to be equal zero, i. e. we neglect thermal evaporation of PD from dislocation loops.

The capture efficiency is defined per unit length of the dislocation line [10]

$$Z = \frac{J}{2\pi R D \overline{C}}, \quad (16)$$

where $J = \oint_S \mathbf{j} d\mathbf{S}$ is the total flux of PD towards the loop.

Following [7,16] we define the new variable

$$\psi(r,z) = \frac{C(r,z)}{\overline{C}} \exp\left[\frac{U(r,z)}{k_B T}\right] \quad (17)$$

and reduce Equation (14) to the form

$$\frac{\partial^2 \psi}{\partial r^2} + \frac{\partial^2 \psi}{\partial z^2} + \left(\frac{1}{r} - \frac{\partial E}{\partial r}\right)\frac{\partial \psi}{\partial r} - \frac{\partial E}{\partial z}\frac{\partial \psi}{\partial z} = 0, \tag{18}$$

where $E(r,z) = U(r,z)/k_B T$. The boundary conditions are

$$\psi = \exp(E) \quad \text{at the external boundary } \sigma \tag{19}$$

$$\psi = 0 \quad \text{at the internal boundary } \tau \tag{20}$$

Additional boundary conditions are specified by zero flux through symmetry lines $o_r$ and $o_z$ (Figure 2), i.e.

$$\partial \psi / \partial r = 0 \text{ at } o_z \tag{21}$$

$$\partial \psi / \partial z = 0 \text{ at } o_r \tag{22}$$

The capture efficiency in terms of the new variable $\psi$ is given by

$$Z = -\frac{1}{2\pi R}\oint_S e^{-E} \nabla \psi d\mathbf{S}, \tag{23}$$

where integration is carried over any closed surface $S$ encapsulating the loop core.

Boundary value problem given by Equations (18)-(22) is solved numerically. To this end Equation (18) is transformed into a set of algebraic equations using the conservative central finite differencing scheme on a rectangular mesh $(i,j)$

$$a_{i,j}^{(1)}\psi_{i+1,j} + a_{i,j}^{(2)}\psi_{i-1,j} + a_{i,j}^{(3)}\psi_{i,j} + a_{i,j}^{(4)}\psi_{i,j+1} + a_{i,j}^{(5)}\psi_{i,j-1} = 0 \tag{24}$$

where $\psi_{i,j} = \psi(r_i, z_j)$. The positions of the mesh point are given by

$$r_i = \begin{cases} i\Delta r & R < L \quad \text{small loop} \\ R - L + i\Delta r & R < L \quad \text{large loop} \end{cases} \quad i = 0,\dots,i_{\max}$$

$$z_j = j\Delta z, \quad j = 0,\dots,j_{\max} \tag{25}$$

The maximum number of points $(i_{\max}, j_{\max})$ along $r$ and $z$ directions are defined by the size of computation domain and mesh spacings $\Delta r$ and $\Delta z$. In Equation (24) coefficients are given by

$$a^{(1)} = \begin{cases} 2 - \dfrac{\partial E}{\partial r}\dfrac{\Delta r}{2}, & r = 0 \\ 1 + \left(\dfrac{1}{r} - \dfrac{\partial E}{\partial r}\right)\dfrac{\Delta r}{2}, & r > 0 \end{cases}, \qquad (26)$$

$$a^{(2)} = \begin{cases} 2 + \dfrac{\partial E}{\partial r}\dfrac{\Delta r}{2}, & r = 0 \\ 1 - \left(\dfrac{1}{r} - \dfrac{\partial E}{\partial r}\right)\dfrac{\Delta r}{2}, & r > 0 \end{cases}, \qquad (27)$$

$$a^{(3)} = \begin{cases} -2\left(2 + \dfrac{\Delta r^2}{\Delta z^2}\right), & r = 0 \\ -2\left(1 + \dfrac{\Delta r^2}{\Delta z^2}\right), & r > 0 \end{cases} \qquad (28)$$

$$a^{(4)} = \dfrac{\Delta r^2}{\Delta z^2} - \dfrac{1}{2}\dfrac{\partial E}{\partial z}\dfrac{\Delta r^2}{\Delta z}, \qquad a^{(5)} = \dfrac{\Delta r^2}{\Delta z^2} + \dfrac{1}{2}\dfrac{\partial E}{\partial z}\dfrac{\Delta r^2}{\Delta z} \qquad (29)$$

Here the subscripts of the coefficients $i$ and $j$ are omitted for brevity.

The dimensionless interaction energy $E(k)$ and its derivatives depend on variable $k$, which is a function of $r_i$ and $z_j$ ($0 \leq k < 1$). To save processing time we builds pre-computed lookup tables of complete elliptic integrals in a discrete set of $k$-points (typically in more that $10^3$ points) using the Legendre normal form. For intermediate values of $k$ we use binary search of lookup tables and linear interpolation. Equation (24) is solved by the successive overrelaxation method (SOR) [8] using an iterative scheme

$$\psi_{i,j}^{(n+1)} = \psi_{i,j}^{(n)} - \dfrac{\gamma}{a_{i,j}^{(3)}} R_{i,j}^{(n+1)}$$

$$R_{i,j}^{(n+1)} = a_{i,j}^{(1)}\psi_{i+1,j}^{(n)} + a_{i,j}^{(2)}\psi_{i-1,j}^{(n+1)} + a_{i,j}^{(3)}\psi_{i,j}^{(n)} + a_{i,j}^{(4)}\psi_{i,j+1}^{(n)} + a_{i,j}^{(5)}\psi_{i,j-1}^{(n+1)} \qquad (30)$$

where $n$ enumerates iterations, $\gamma$ is the overrelaxation parameter and $R_{i,j}^{(n+1)}$ is the residual vector. At $\gamma = 1$ Equation (30) is equivalent to Equation (24). On a rectangular $i_{\max} \times j_{\max}$ mesh SOR converges most rapidly when $\gamma$ is assigned the optimum value [18]

$$\gamma = \dfrac{2}{1 + \sqrt{1 - \chi^2}}, \qquad \chi = \dfrac{\cos(\pi/i_{\max}) + (\Delta r/\Delta z)^2 \cos(\pi/j_{\max})}{1 + (\Delta r/\Delta z)^2} \qquad (31)$$

In our problem the integration domain is not rectangular; however, the value of the overrelaxation parameter given by Equation (31) produces a good convergence of the iterative algorithm.

The mesh is scanned in order of increasing $i$ and $j$. The sequence of computation with Equation (30) starts from the lower left corner, proceeds upward until reaching the top boundary, and then goes to the bottom of the next vertical line on the right. This process is repeated until the new value of $\psi$ at the last interior point at the upper right corner has been obtained. The newly calculated values of $\psi_{i,j}$ are used automatically in later calculations during the next scan of the mesh.

According to [7] the condition for the finite difference solution to be stable is that coefficients in the residual vector change slowly, i.e. the mesh spacings, $\Delta r$ and $\Delta z$, have to be small

$$\left| \frac{1}{r} - \frac{\partial E}{\partial r} \right| \frac{\Delta r}{2} < 1, \qquad \frac{\partial E}{\partial z} \frac{\Delta z}{2} < 1 \tag{32}$$

These conditions impose restrictions on the choice of the radius of the dislocation core, near which the interaction energy exhibits high gradients. In our calculations we use equal mesh spacings $\Delta r$ and $\Delta z$ along $r$ and $z$ axis.

The initial guess to the solution $\psi^{(0)}$ is zero in all interior mesh points, in addition to those prescribed at the boundary $\sigma$ and $\tau$ of the influence region (Figure 2). To satisfy boundary conditions at the symmetry lines $o_r$ and $o_z$ (Figure 2), i.e. zero flux across the symmetry lines, we introduce ghost points $\psi^{(n)}_{-1,j} = \psi^{(n)}_{1,j}$ and $\psi^{(n)}_{i,-1} = \psi^{(n)}_{i,1}$.

The norm of the residual vector $R^{(n+1)}_{i,j}$ is used as a criterion for terminating the iteration. An arbitrary surface $S$ in the form of a rectangle of rotation is chosen in the region of influence to calculate the capture efficiency given by Equation (23).

**Results of calculations**

The parameters typical for an alloy based on Al-Pd-Mn were used in calculations: The temperature $T = 600$ K, the shear modulus $\mu = 70$ GPa, the Poisson ratio $\nu = 0.27$, the average atomic volume $\omega = 0.015$ nm$^3$, the relaxation volumes of vacancies $\Omega_v = -0.5\omega$ and interstitial atoms $\Omega_i = \omega$, the dislocation core radius $2b$, the mesh spacing $\Delta r = \Delta z = 0.2b$. Calculations are performed for phason rings of the width of $25b$, $50b$ and $150b$. If the loop radius is less than the width of the ring, then the ring fills the loop interior. The relative tolerance of the calculations is $10^{-4}$.

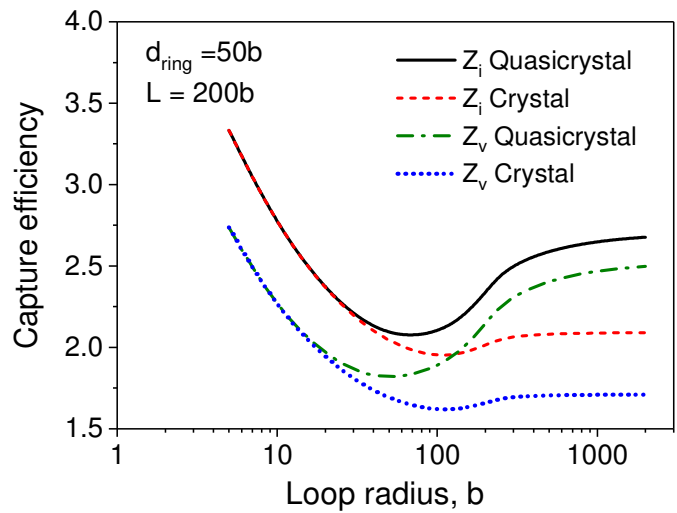

Figure 3. *Capture efficiencies of dislocation loop in crystal and quasicrystal. The loop radius and the size of influence region are given in units of the Burgers vector.*

Figure 3 compares the dislocation capture efficiencies for vacancies and interstitial atoms by a dislocation loops with and without the phason ring. In the region of small sizes of loops, the capture efficiency does not depend on the type of material. This is due to the fact that, for small loops the effective radius of elastic interaction with PD exceeds the radius of the dislocation loop. That is, such loops behave like a 3D sink, the internal structure of which does not affect the total PD flux. At larger sizes the capture efficiency of loops with phason rings is higher than that of loops without rings. With a further increase of the loop radius, the capture efficiency reaches saturation, i.e. tends to the capture efficiency of a straight edge dislocation.

Figure 4 shows that the capture efficiency depends on mean distance between sinks, i.e., on sink density or, in terms of the effective medium, on total sink strength. Figure 5 shows the dependence of the bias factor $B_L = 1 - Z_{Lv}/Z_{Li}$ on loop radius. It is seen that the bias factor of the dislocation loop with a phason ring is smaller than the bias factor of the loop in a crystal. Moreover, the bias factor significantly depends on the width of phason ring. At the same time, the distance between sinks does not strongly affect the dependence of the bias factor on loop radius (Figure 6).

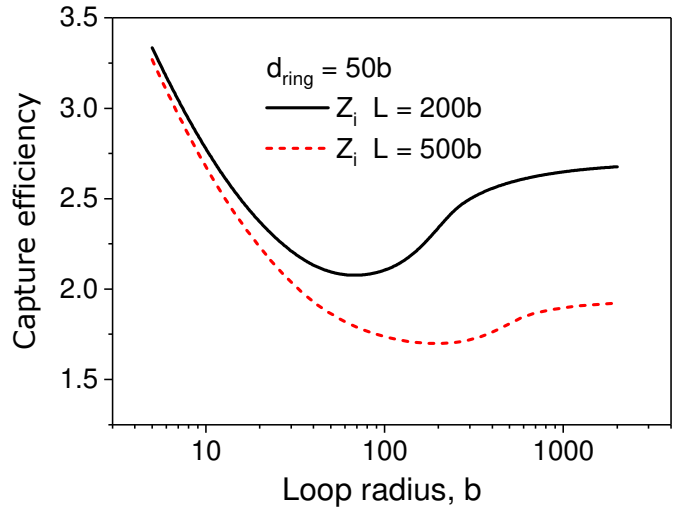

Figure 4. *Radius dependence of capture efficiency of interstitial atoms by a loop. Calculations were carried out for two sizes of influence region which are indicated in the graph.*

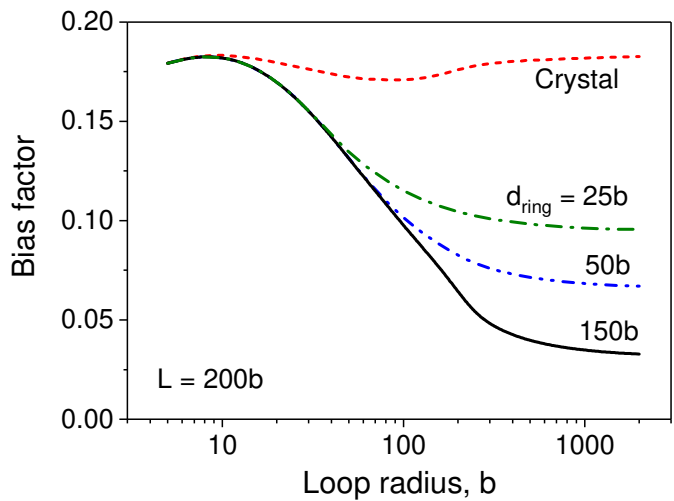

Figure 5. *Dependence of the bias factor on loop radius in a crystal and a quasicrystal. Widths of the phason ring are d = 25b, 50b and 150b.*

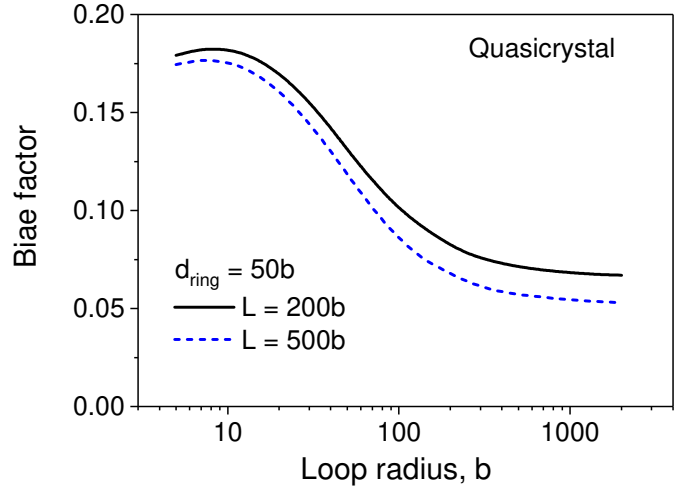

Figure 6. *Dependence of the bias factor on loop radius and size of influence region L = 200b and 500b.*

According to our calculations, the bias factor of dislocation loops in quasicrystals is smaller than that in a crystalline material. Therefore, at the same initial conditions and the same distribution function of dislocation loops, the growth rate of vacancy voids and, accordingly, the swelling of quasicrystals are predicted to be low compared to crystalline materials. Using Equation (13), we estimate $S'_{Quasi}/S'_{Cryst}$, the decrease of the swelling rate of a quasicrystal compared to the swelling rate of a crystal with the same microstructural parameters listed in Table 1.

Table 1. Parameters for evaluation of a decrease of the quasicrystal swelling rate.

| Parameter | Value |
| --- | --- |
| Burger's vector, $b$, nm | 0.25 |
| Fraction of freely-migrating PD, $\eta$ | 0.25 |
| Density of straight edge dislocations, $\rho_{str}$, m$^{-2}$. | $10^{14}$ |
| Sink strength of edge dislocations, $Z_{str,v}\rho_{str}$, m$^{-2}$ | $1.7 \times 10^{14}$ |
| Number density of loop, $n_L$, m$^{-3}$ | $10^{21}$ |
| Sink strength of loops, $k_{Lv}^2$, m$^{-2}$ | $2 \times 10^{14}$ |
| Sink strength of voids, $k_C^2$, m$^{-2}$ | $3.6 \times 10^{14}$ |

The size distribution of loops $f(r)$ is selected in the form of the Gaussian distribution (Figure 7)

$$f(r) = 3.4 \times 10^{28} \text{m}^{-4} \exp\left(-\frac{(r-15\text{nm})^2}{400\text{nm}^2}\right). \tag{33}$$

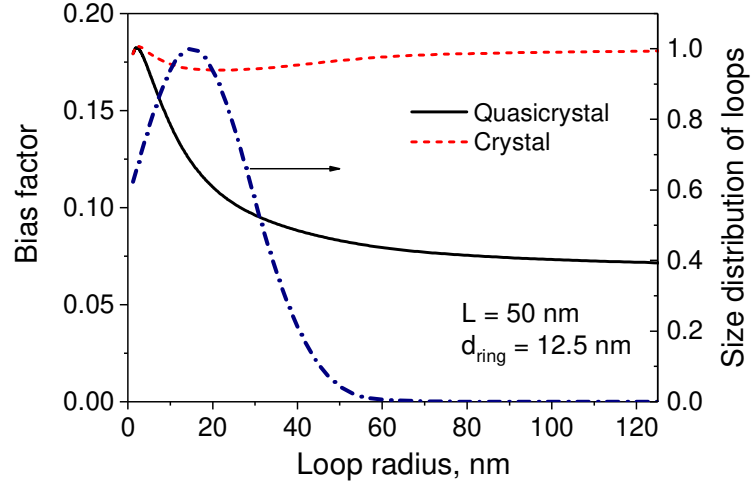

Figure 7. *Loop bias factor and the size distribution (33) (without the pre-exponential factor).*

We consider an early stage of void formation (the transition stage or incubation period) and a later stage, when the swelling rate of many reactor steels and alloys tends to saturation $S'_{max} \sim 1\%/\text{dpa}$ [19-21]. Temperature is assumed to be sufficiently high, when PD are predominantly absorbed by sinks, and the bulk recombination of PD can be neglected.

At the early stage of void formation, the main sinks for PD are straight dislocations and large interstitial dislocation loops (in the model of this paper these are the largest loops $r = 2000b$ with the corresponding capture efficiencies)

$$Z_{str,n}\rho_{str} \gg k_C^2 + 2\pi \int Z_n(r) r f(r) dr . \tag{34}$$

At the early stage the swelling rate is low

$$S' = \eta B_{mean} \frac{k_C^2}{Z_{str,v}\rho_{str}} \approx \eta B_{str} \frac{k_C^2}{Z_{str,v}\rho_{str}} \ll \eta B_{str} . \tag{35}$$

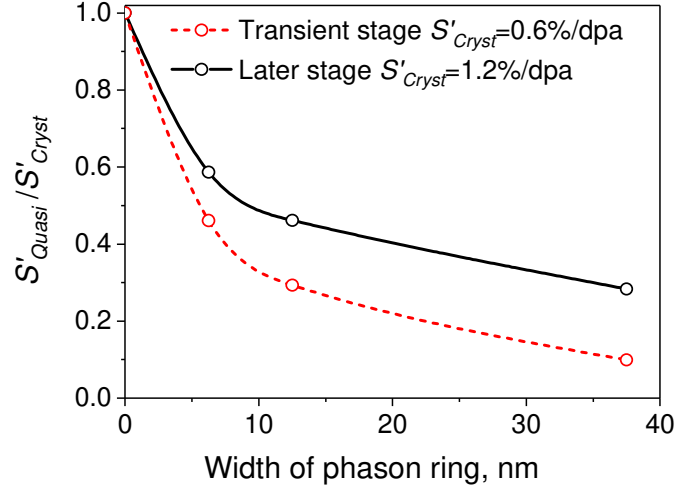

Figure 8. *The effect of the phason ring width on the reduction of the swelling rate in the quasicrystal compared to the crystalline material. The dotted line corresponds to the transient stage of irradiation; the sink strengths of voids and dislocations are about 10% of sink strengths at the later stage (see Table 1). The solid line is the ratio of the maximum swelling rates of the quasicrystal and the crystalline material.*

Figure 8 demonstrates that at the early stage (dashed line) $S'_{Quasi}/S'_{Cryst} = B_{str}^{Quasi}/B_{str}^{Cryst} < 1$. This means that the duration of the incubation period for void swelling of quasicrystals is larger than that of crystalline materials.

The maximum swelling rate is achieved at a later stage, when sink strengths of voids and dislocations are approximately equal $k_{dv}^2 \approx k_C^2$

$$S' = \eta B_{\text{mean}} \frac{k_C^2}{k_{dv}^2 + k_C^2} \sim 0.5 \eta B_{\text{mean}}. \quad (36)$$

Figure 8 shows the ratio of the maximum swelling rates of the quasicrystal and the crystalline material. It can be seen that, for all sizes of the phason ring, the swelling rate of the quasicrystal is less than the swelling rate of the crystalline material with similar microstructural parameters

## Conclusions

The radiation-induced vacancy swelling of quasicrystals was considered. Similar to crystalline materials, in quasicrystals dislocations preferentially absorb interstitial atoms which results in vacancy swelling of materials. In quasicrystals the growth of dislocations is accompanied by formation of phasons – localized topological defects of vacancy and interstitial types. Since the diffusion mobility of phasons is much lower than that of vacancies, at relatively low temperatures phasons form a ring-shaped trail inside a growing interstitial dislocation loop. The capture efficiencies of a dislocation loop with a complementary phason ring for PD were found by the SOR numerical method. We have shown that the phason ring significantly reduces the dislocation bias to interstitial atoms. For this reason quasicrystalline materials are predicted to demonstrate an increased resistance to vacancy swelling. At elevated temperatures phasons become mobile, and the phason rings dissolve. Nevertheless, a significant reduction of the vacancy swelling is expected, since mobile phasons dispersed in the quasicrystal serve as recombination centers for radiation-induced vacancies and interstitial atoms.

In summary, the generation of phasons by evolving dislocations leads to suppression of radiation-induced vacancy swelling of quasicrystals.

## Disclosure statement

No potential conflict of interest was reported by the authors.